\documentclass[aps,twocolumn,superscriptaddress]{revtex4}

\usepackage{graphicx}

\newcommand{\bphi}{{\mbox{\boldmath $\varphi$}}}
\newcommand{\bomega}{{\mbox{\boldmath $\omega$}}}
\newcommand{\bdelta}{{\mbox{\boldmath $\delta$}}}

\begin{document}

\title{Non-additivity of decoherence rates in superconducting qubits}

\author{Guido Burkard}
\affiliation{IBM T.\ J.\ Watson Research Center,
         P.\ O.\ Box 218,
         Yorktown Heights, NY 10598, USA}
\author{Frederico Brito}
\affiliation{Departamento de F\'{\i}sica da Mat\'eria Condensada,
Instituto de F\'{\i}sica Gleb Wataghin,
Universidade Estadual de Campinas,
Campinas-SP 13083-970, Brazil}
\affiliation{IBM T.\ J.\ Watson Research Center,
         P.\ O.\ Box 218,
         Yorktown Heights, NY 10598, USA}

\begin{abstract}
We show that the relaxation and decoherence rates $T_1^{-1}$ and $T_2^{-1}$ of a qubit coupled to
several noise sources are in general not additive, i.e., that the total rates are not 
the sums of the rates due to each individual noise source.
To demonstrate this, we calculate the relaxation and pure dephasing rates 
$T_1^{-1}$ and $T_\phi^{-1}$ of a superconducting (SC) flux qubit in the Born-Markov approximation in the presence of
several circuit impedances $Z_i$ using network graph theory and determine
their deviation from additivity (the mixing term).
We find that there is no mixing term in $T_\phi^{-1}$ and that the mixing terms in $T_{1}^{-1}$ and $T_2^{-1}$
can be positive or negative, leading to reduced or enhanced relaxation and decoherence times $T_1$ and $T_2$.  
The mixing term due to the circuit inductance $L$ at the qubit transition
frequency $\omega_{01}$ is generally of second order in $\omega_{01}L/Z_i$, 
but of third order if all impedances $Z_i$ are pure resistances.
We calculate $T_{1,2}$ for an example of a SC flux qubit coupled to two impedances.
\end{abstract}

\maketitle

\textit{Introduction.}
The loss of quantum coherence and the transition from quantum to classical
behavior has been a long-standing fundamental problem 
\cite{CaldeiraLeggett,Zurek}.
More recently,
the phenomenon of decoherence has attracted much interest in a new context,
because quantum coherence is an essential prerequisite for quantum computation.
For some systems that have been proposed as physical realizations
of quantum hardware (see, e.g., Ref.~\onlinecite{Fortschritte}), there have been 
extensive studies, both in theory and experiment, of the mechanisms that are causing 
decoherence. 
Generally, an open quantum system loses coherence by interacting with a large number 
of external degrees of freedom (heat bath, environment).  It is the physical
nature of the environment and the system-environment coupling that distinguishes
the various mechanisms of decoherence.  It is quite natural that for a given
open quantum system there will be \textit{several} distinct decoherence mechanisms.
Previous studies have typically tried to identify the strongest source of 
decoherence, i.e., the one that leads to the shortest relaxation and decoherence times, $T_1$ and $T_2$,
and to analyze the corresponding mechanism in order to predict
decoherence times.  In the presence of several decoherence sources for the same
system, the decoherence rates $T_1^{-1}$ and $T_2^{-1}$ have usually been quoted separately 
for each source.
Often, it is assumed that the total decoherence or relaxation rate is the sum of the rates
corresponding to the various sources (see, e.g., Ref.~\onlinecite{MSS} for
the case of superconducting qubits).  In the theory of electron scattering in
metals, this assumption is also known as Matthiessen's rule \cite{AM}.
In this paper, we show that the total decoherence and relaxation rates of a quantum system
in the presence of several decoherence sources are \textit{not} necessarily the 
sums of the rates due to each of the mechanisms separately,
and that the corrections to additivity (mixing terms) can have both signs.

We investigate the decoherence due to several sources in superconducting (SC) 
flux qubits \cite{Mooij,Orlando,vanderWal,Chiorescu,Friedman,IBM} 
(see Ref.~\onlinecite{MSS} for a review of SC qubits);
the general idea of the present analysis
may however be applied to other systems as well.
SC flux qubits are small SC circuits that contain
Josephson junctions.  The differences $\varphi_i$ of the SC phases
across the junctions $J_i$, where $i=1,\ldots,n$, are the relevant quantum degrees 
of freedom of the
system;  we denote the quantum operator of these phase differences 
collectively with the
vector $\bphi=(\varphi_1,\varphi_2,\ldots,\varphi_n)$.   
The circuit is constructed such that it gives rise to a potential $U(\bphi)$
which forms a double well and therefore can be used to encode one qubit.   
In our analysis, we will make
use of a recently developed circuit theory describing the dissipative dynamics 
of arbitrary SC flux qubits \cite{BKD}.
Our analysis relies on the theory for open
quantum systems introduced by Caldeira and Leggett \cite{CaldeiraLeggett} where the
dissipative elements (impedances $Z_i$) are represented by a set of baths of harmonic oscillators
(an alternative approach to a quantum theory of dissipative electric circuits is 
to represent impedances as infinite transmission lines \cite{Yurke84}).
\begin{figure}[hb]
\centerline{\includegraphics[width=7cm]{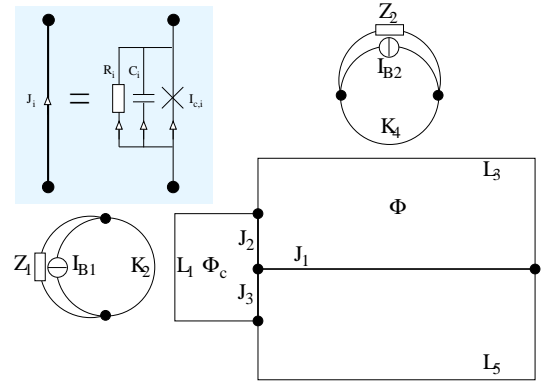}}
\caption{\label{circuit}
Circuit graph of the gradiometer qubit \cite{IBM-unpub}, under the influence
of noise from two sources $Z_1$ and $Z_2$.
Branches of the graph denote Josephson junctions $J_i$,
inductances $L_i$ and $K_i$, current sources $I_{Bi}$,
and external impedances $Z_i$, and are connected by the nodes (black dots) of the
graph.
Inset:  A resistively-shunted Josephson junction (RSJ) $J_i$, represented by a thick 
line in the circuit graph, is modeled by an ideal junction (cross) with critical current 
$I_{ci}$, shunt resistance $R_i$, and junction capacitance $C_i$.}
\end{figure}

For concreteness, we
demonstrate our theory on the example of the gradiometer qubit with $n=3$ junctions that
is currently under experimental investigation \cite{IBM-unpub}, see Fig.~\ref{circuit}.
We emphasize, however, that our findings are completely general and apply to arbitrary SC flux qubits.
The qubit is controlled by applying a magnetic flux $\Phi_c$ to the small loop
on the left by driving a current $I_{B1}$ in a coil next to it, and simultaneously 
by applying a magnetic flux $\Phi$ on one side of the gradiometer using $I_{B2}$.
Real current sources are not ideal, i.e., they are characterized by a finite
frequency-dependent impedance $Z_i(\omega)$, giving rise to decoherence
of the qubit \cite{Tian99,Tian02,WWHM,Wilhelm}.  Since the shunt resistances
$R_i$ of the junctions are typically much larger ($>{\rm M}\Omega$) than the impedances
of the current sources (between $\approx 50\,\Omega$ and $\approx 10\,{\rm k}\Omega$), we
concentrate in our example on the impedances $Z_1$ and $Z_2$
of the two current sources.

Using circuit graph theory \cite{BKD}, 
we obtain the classical equations of motion 
of a general SC circuit in the form
\begin{equation}
  \label{eq-mot}
  {\bf C} \ddot {\bf \bphi} = -\frac{\partial U}{\partial\bphi} - {\bf M}*\bphi ,
\end{equation}
where ${\bf C}$ is the $n\times n$ capacitance matrix and $U(\bphi; I_{B1}, I_{B2})$ 
is the potential.  The dissipation matrix ${\bf M}(t)$ is a real, symmetric, and causal 
$n\times n$ matrix, i.e., ${\bf M}(t)^T = {\bf M}(t)$ for all $t$, and ${\bf M}(t)=0$ 
for $t<0$.  
The convolution is defined as $(f*g)(t) = \int_{-\infty}^t f(t-\tau)g(\tau)d\tau$.
Since it is not explicitly used here, we will not further specify $U$.
The dissipation matrix in the Fourier representation \cite{footnote1},
${\bf M}(\omega) = \int_0^\infty e^{-i\omega t-\epsilon t} {\bf M}(t) \, dt$,
where $\epsilon>0$ has been introduced to ensure convergence (at the end,
$\epsilon\rightarrow 0$), can be found from circuit theory \cite{BKD} as
\begin{equation}
  \label{Mcircuit}
  {\bf M}(\omega) = \bar{\bf m}\bar{\bf L}_Z(\omega)^{-1}\bar{\bf m}^T,
\end{equation}
where $\bar{\bf m}$ denotes a real $n\times n_Z$ matrix that can be obtained from the
circuit inductances, and
where the $n_Z\times n_Z$ matrix $\bar{\bf L}_Z(\omega)$ has the form
\begin{equation}
  \label{LZba}
  \bar{\bf L}_Z(\omega)  =  {\bf L}_Z(\omega)  +  {\bf L}_{\rm c}.
\end{equation}
Here, $n_Z$ is the number of impedances in the circuit
(in our example, $n_Z=2$) and ${\bf L}_Z(\omega)= {\bf Z}(\omega)/i\omega$,
where ${\bf Z}(\omega)$ the impedance matrix.
The frequency-independent and real inductance matrix
${\bf L}_{\rm c}$ can be obtained from the circuit inductances \cite{BKD}.
Since we start from independent
impedances, ${\bf Z}$ and ${\bf L}_Z$ are diagonal.
Moreover, note that
\begin{equation}
\label{positivity}
{\rm Im}\bar{\bf L}_Z^{-1}
=\omega \!\! \left[{\rm Re}{\bf Z}(\omega)
+\omega^2 \tilde{\bf L}_{\rm c}(\omega)\left({\rm Re}{\bf Z}(\omega)\right)^{-1}\tilde{\bf L}_{\rm c}(\omega)\right]^{-1}\!\!\!\!\!\!,
\end{equation}
where $\tilde{\bf L}_{\rm c}(\omega)={\bf L}_{\rm c}+{\rm Im}{\bf Z}(\omega)/\omega$,
thus it follows from ${\rm Re}{\bf Z}>0$ that 
${\rm Im}\bar{\bf L}_Z^{-1}$ and ${\rm Im}{\bf M}$ are positive matrices.

\textit{Multi-dimensional Caldeira-Leggett model.}
We now construct a Caldeira-Leggett Hamiltonian \cite{CaldeiraLeggett},
${\cal H} = {\cal H}_S + {\cal H}_B + {\cal H}_{SB}$,
that reproduces the classical dissipative equation of motion, Eq.~(\ref{eq-mot}),
and that is composed of parts for the system (S), for $m \ge 1$ harmonic oscillator baths (B), 
and for the system-bath (SB) coupling,
\begin{eqnarray}
  {\cal H}_S &=& \frac{1}{2}{\bf Q}^T{\bf C}^{-1}{\bf Q} 
                 + \left(\frac{\Phi_0}{2\pi}\right)^2 U(\bphi),\label{HS}\\
  {\cal H}_B &=& \sum_{j=1}^m \sum_\alpha \left(\frac{p_{\alpha j}^2}{2 m_{\alpha j}}+\frac{1}{2}m_{\alpha j} \omega_{\alpha j}^2 x_{\alpha j}^2\right),\label{HB}\\
  {\cal H}_{SB} &=&  \sum_\alpha \bphi ^T {\bf c}_\alpha {\bf x}_\alpha , \label{HSB}
\end{eqnarray}
where the capacitor charges ${\bf Q}$ are the canonically conjugate momenta
corresponding to the Josephson fluxes $(\Phi_0/2\pi)\bphi$, 
where ${\bf x}_\alpha = (x_{\alpha 1}, \ldots, x_{\alpha m})$,
and ${\bf c}_\alpha$ is a real $n\times m$ matrix.
From the classical equations of motion of the system and bath coordinates and by
taking the Fourier transform, we obtain Eq.~(\ref{eq-mot}),  with
${\bf M}(\omega) = (2\pi/\Phi_0)^2\sum_\alpha {\bf c}_\alpha [{\bf m}_{\alpha}
(\omega^2-{\bomega}_{\alpha}^2)]^{-1} {\bf c}_\alpha^T = {\bf M}(\omega)^T$,
where the $m\times m$ mass and frequency matrices ${\bf m}_{\alpha}$ and ${\bomega}_{\alpha}$ are diagonal
with entries $m_{\alpha j}$ and $\omega_{\alpha j}$.  
Using the regularization $\omega\rightarrow \omega-i\epsilon$ when taking
Fourier transforms also guarantees that ${\bf M}(t)$ is causal and real.

Defining the spectral density of the environment as the matrix function
\begin{equation}
  \label{Jdef}
  {\bf J}(\omega) = \frac{\pi}{2}\sum_\alpha {\bf c}_\alpha{\bf m}_{\alpha}^{-1}{\bomega}_{\alpha}^{-1}
                                                \bdelta({\omega-\bomega_\alpha}){\bf c}_\alpha^T,
\end{equation}
where $\bdelta_{ij}({\bf X})\equiv \delta({\bf X}_{ij})$, we find the relation
\begin{equation}
  \label{J}
  {\bf J}(\omega) = \left(\frac{\Phi_0}{2\pi}\right)^2\!\!{\rm Im}{\bf M}(\omega) 
                  = \sum_{j=1}^m J_j(\omega) {\bf m}_j(\omega){\bf m}_j(\omega)^T,
\end{equation}
where we have used the
spectral decomposition of the real, positive, and symmetric matrix \cite{footnote1}
${\rm Im} {\bf M}(\omega)$,
with the eigenvalues $J_j(\omega)>0$ and the real and normalized
eigenvectors ${\bf m}_j(\omega)$.  The integer $m\leq n,n_Z$ denotes the 
maximal rank of ${\rm Im}{\bf M}(\omega)$, i.e., $m=\max_\omega \left({\rm rank} \left[ {\rm Im}{\bf M}(\omega)\right]\right)$.
Using Eq.~(\ref{J}), and choosing $c_{\alpha ij} = \gamma_{\alpha j}{\bf m}_i(\omega_{\alpha j})$,
we find that $J_j(\omega)$ is the spectral density of the $j$-th
bath of harmonic oscillators in the environment,
$J_j(\omega) = (\pi/2)\sum_\alpha (\gamma_{\alpha j}^2/m_{\alpha j}\omega_{\alpha j})
\delta(\omega-\omega_{\alpha j})$.

The master equation of the reduced system density matrix $\rho_S={\rm Tr}_B\rho$ in the 
Born-Markov approximation, expressed in the eigenbasis $\{|m\rangle\}$ of ${\cal H}_S$,
yields the Bloch-Redfield equation \cite{Redfield},
$\dot{\rho}_{nm}(t) = -i\omega_{nm}\rho_{nm}(t) -\sum_{kl}R_{nmkl}\rho_{kl}(t)$,
where ${\rho}_{nm}=\langle n|\rho_S|m\rangle$, $\omega_{nm}=\omega_n-\omega_m$,
and $\omega_m$ is the eigenenergy of ${\cal H}_S$ corresponding to the eigenstate $|m\rangle$.
The Redfield tensor has the form
$R_{nmkl} = \delta_{lm}\!\sum_r \Gamma_{nrrk}^{(+)} + \delta_{nk}\!\sum_r \Gamma_{lrrm}^{(-)}
-\Gamma_{lmnk}^{(+)}-\Gamma_{lmnk}^{(-)}$,
with the rates
$\Gamma_{lmnk}^{(+)} =  \int_0^\infty dt \exp(-it\omega_{nk}){\rm Tr}_B \tilde{\cal H}_{SB}(t)_{lm}\tilde{\cal H}_{SB}(0)_{nk}\rho_B$
and  $(\Gamma_{knml}^{(-)})^*=\Gamma_{lmnk}^{(+)}$, where
$\tilde{\cal H}_{SB}(t)_{nm} = \langle n| e^{it{\cal H}_B}{\cal H}_{SB}e^{-it{\cal H}_B}|m\rangle$.
For the system-bath interaction Hamiltonian, Eq.~(\ref{HSB}), we obtain
\begin{eqnarray}
{\rm Re}\Gamma_{lmnk}^{(+)} &=& \bphi_{lm}^T {\bf J}(|\omega_{nk}|) \bphi_{nk} \frac{e^{-\beta\omega_{nk}/2}}{\sinh(\beta |\omega_{nk}|/2)},\label{ReGamma}\\
{\rm Im}\Gamma_{lmnk}^{(+)} &=& -\frac{2}{\pi}P\!\!\int_0^\infty 
                                 \frac{\bphi_{lm}^T {\bf J}(\omega)\bphi_{nk}}{\omega^2-\omega_{nk}^2}
                                 \left(\omega-\omega_{nk}\coth\frac{\beta\omega}{2}\right),\nonumber
\end{eqnarray}
where $\bphi_{nk} = \langle n|\bphi|k\rangle$.
For two levels $n=0,1$, and within the secular approximation, we can determine the relaxation
and decoherence rates $T_1^{-1}$ and $T_2^{-1}$ in the Bloch equation as \cite{BKD}
$T_1^{-1}    = 2{\rm Re}(\Gamma_{0110}^{(+)}+\Gamma_{1001}^{(+)})$
and $T_2^{-1}    = (2 T_1)^{-1} +  T_\phi^{-1}$,
where $T_\phi^{-1} = {\rm Re}(\Gamma_{0000}^{(+)}+\Gamma_{1111}^{(+)}-2\Gamma_{0011}^{(+)})$
is the pure dephasing rate.
Using Eq.~(\ref{ReGamma}), we find
\begin{eqnarray}
  T_1^{-1} &=& 4 \bphi_{01}^\dagger {\bf J}(\omega_{01})\bphi_{01} \coth\left(\frac{\beta\omega_{01}}{2}\right),\label{T1}\\
  T_\phi^{-1} &=& \frac{2}{\beta}\lim_{\omega\rightarrow 0} (\bphi_{00}-\bphi_{11})^\dagger\frac{{\bf J}(\omega)}{\omega}(\bphi_{00}-\bphi_{11}).\label{Tphi}
\end{eqnarray}
With the spectral decomposition, Eq.~(\ref{J}), we obtain
\begin{eqnarray}
    T_1^{-1} &=& 4 \sum_{j=1}^m |\bphi_{01}\!\cdot {\bf m}_j(\omega_{01})|^2 J_j(\omega_{01}) \coth\left(\frac{\beta\omega_{01}}{2}\right)\!,\quad\quad \label{T1s}\\
  T_\phi^{-1} &=& \frac{2}{\beta} \sum_{j=1}^m  |{\bf m}_j(0)\cdot (\bphi_{00}-\bphi_{11})|^2  \left.\frac{J_j(\omega)}{\omega}\right|_{\omega\rightarrow 0}.\label{Tphis}
\end{eqnarray}
In the last equation, we have used that the limit ${\bf m}_j(0)=\lim_{\omega\rightarrow 0}{\bf m}_j(\omega)$
exists because $|{\bf m}_j(\omega)|^2=1$ and thus all components of ${\bf m}_j(\omega)$ are bounded.

\textit{Mixing Terms.}
\begin{figure}
\centerline{\includegraphics[width=8.5cm]{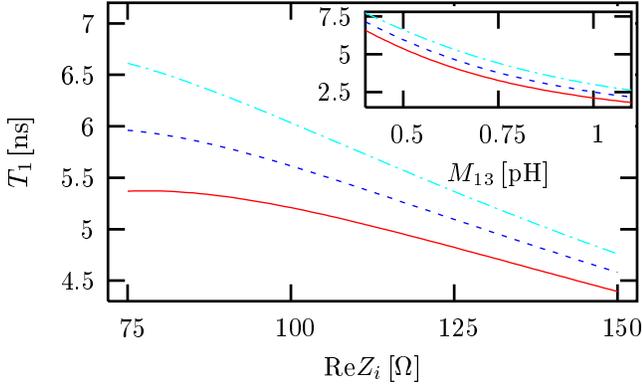}}
\caption{\label{fig2}
The relaxation rate $T_1$ without the mixing term (dashed blue line),
and including the mixing term for $R_{\rm im}=+10\,{\rm k}\Omega$ (solid red line)
and $R_{\rm im}=-10\,{\rm k}\Omega$ (dot-dashed light blue line), 
for $M_{13}=0.5\,{\rm pH}$ as a function of ${\rm Re}Z_i$.
Inset:  $T_1$ for $R={\rm Re Z}_i=75\,\Omega$
for a range of mutual inductances $M_{13}$.}
\end{figure}
In the case where ${\bf L}_{\rm c}$ is diagonal, or if its off-diagonal elements
can be neglected because they are much smaller than ${\bf L}_Z(\omega)$ for
all frequencies $\omega$,
we find, using Eq.~(\ref{LZba}), that the contributions due to different 
impedances $Z_i$ are independent, thus $m=n_Z$ and
${\bf M}(\omega) = \bar{\bf m}\bar{\bf L}_Z(\omega)^{-1}\bar{\bf m}^T 
= \sum_j \bar{\bf m}_j \bar{\bf m}_j^T i\omega/(Z_j(\omega)+i\omega L_{jj})$,
where ${\bf m}_j = \bar{\bf m}_j$ is simply the $j$-th column of the matrix $\bar{\bf m}$
and $L_{jj}$ is the $j$-th diagonal entry of ${\bf L}_{\rm c}$.
As a consequence, the total rates $1/T_1$ and $1/T_\phi$ are the sums of the individual rates,
$1/T_1^{(j)}$ and $1/T_\phi^{(j)}$, where
\begin{eqnarray}
    \frac{1}{T_1^{(j)}} &=& 4 \left(\frac{\Phi_0}{2\pi}\right)^2 \!\! |\bphi_{01}\cdot \bar{\bf m}_j|^2 {\rm Re}\frac{\omega_{01}\coth\left(\beta\omega_{01}/2\right)}{Z_j(\omega_{01})+i\omega_{01} L_{jj}}, \quad\quad\label{T1-i}\\
    \frac{1}{T_\phi^{(j)}} &=& \frac{2}{\beta}\left(\frac{\Phi_0}{2\pi}\right)^2 |\bar{\bf m}_j\cdot (\bphi_{00}-\bphi_{11})|^2 {\rm Re}\frac{1}{Z_j(0)}. \label{Tphi-i}
\end{eqnarray}
In general, the situation is more complicated because current fluctuations
due to different impedances are mixed by the presence of the circuit.
In the regime ${\bf L}_{\rm c} \ll {\bf L}_Z(\omega)$, we can use Eq.~(\ref{LZba})
to make the expansion
\begin{equation}
  \label{series1}
  \bar{\bf L}_Z^{-1} 
          = {\bf L}_Z^{-1} - {\bf L}_Z^{-1}{\bf L}_{\rm c}{\bf L}_Z^{-1}
                           + {\bf L}_Z^{-1}{\bf L}_{\rm c}{\bf L}_Z^{-1}{\bf L}_{\rm c}{\bf L}_Z^{-1}
                           - \cdots \,.
\end{equation}
The series Eq.~(\ref{series1}) can be partially resummed,
\begin{equation}
  \label{partialresum}
  \bar{\bf L}_Z^{-1}(\omega) = {\rm diag}\left(\frac{i\omega}{Z_j(\omega) + i\omega L_{jj}} \right) 
                               + {\bf L}_{\rm mix}^{-1}(\omega).
\end{equation}
The first term in Eq.~(\ref{partialresum}) simply gives rise to the sum of the individual 
rates, as in Eqs.~(\ref{T1-i}) and (\ref{Tphi-i}), while the second term
gives rise to mixed terms in the total rates.   
The rates can therefore be decomposed as ($X=1,2,\phi$)
\begin{equation}
  \label{Tdec}
  \frac{1}{T_X}    = \sum_j \frac{1}{T_X^{(j)}} + \frac{1}{T_X^{\rm (mix)}}.
\end{equation}
For the mixing term in the relaxation rate, we find
\begin{equation}
  \frac{1}{T_1^{\rm (mix)}} = 4 \!\! \left(\frac{\Phi_0}{2\pi}\right)^2\!\!\!\bphi_{01}^\dagger \bar{\bf m}{\rm Im}{\bf L}_{\rm mix}^{-1}(\omega_{01})\bar{\bf m}^T\bphi_{01} \coth\!\!\left(\frac{\beta\omega_{01}}{2}\right).\label{T1-mix}
\end{equation}
We can show that there is no mixing term in the pure dephasing rate, i.e., $1/T_\phi^{\rm (mix)} = 0$,
and consequently, $T_2^{\rm (mix)} = 2T_1^{\rm (mix)}$.
The absence of a mixing term in $T_\phi$ can be understood as follows. Since the first term in
Eq.~(\ref{series1}) only contributes to the first term in Eq.~(\ref{partialresum}),
the low-frequency asymptotic of ${\rm Im}{\bf L}_{\rm mix}(\omega)^{-1}$ involves only $\omega^2$
and higher powers of $\omega$ (it can be assumed that $Z_i(\omega=0)$ is finite),
thus Eq.~(\ref{Tphi}) yields zero in the limit $\omega\rightarrow 0$.
While ${\rm Im}\bar{\bf L}_Z^{-1}$ is a positive matrix, ${\rm Im}{\bf L}_{\rm mix}^{-1}$ does not need to be
positive, therefore the mixing term $1/T_{1}^{\rm mix}$ can be both positive or negative.
Furthermore, we can show that if ${\bf Z}(\omega)$ is real, 
only odd powers of $\omega{\bf L}_{\rm c}{\bf Z}^{-1}$ occur, and in particular, that in this case
${\rm Im}{\bf L}_{\rm mix}(\omega)^{-1} = O(\omega^3)$,
by using Eq.~(\ref{positivity}) to write ${\bf J}(\omega) \simeq  \omega {\bf Z}(\omega)^{-1}
- \omega^3 {\bf Z}(\omega)^{-1}{\bf L}_{\rm c}{\bf Z}(\omega)^{-1}{\bf L}_{\rm c}{\bf Z}(\omega)^{-1}$,
up to higher orders in $\omega{\bf L}_{\rm c}{\bf Z}(\omega)^{-1}$.

In the case of two external impedances, $n_Z=2$,
we can completely resum Eq.~(\ref{series1}), with the result
\begin{widetext}
\begin{equation}
  \label{Lmix-grad}
  {\bf L}_{\rm mix}^{-1}(\omega) = \frac{L_{12}}{(Z_1(\omega)/i\omega+L_{11})(Z_2(\omega)/i\omega+L_{22})-L_{12}^2}
                           \left(\begin{array}{c c}
                               \frac{L_{12}}{Z_1(\omega)/i\omega+L_{11}}  &  -1\\
                               -1                           &  \frac{L_{12}}{Z_2(\omega)/i\omega+L_{22}}
                           \end{array}\right)
                         \approx  -\frac{\omega^2 L_{12}}{Z_1(\omega) Z_2(\omega)} \sigma_x ,
\end{equation}
\end{widetext}
where $L_{ij}$ are the matrix elements of ${\bf L}_{\rm c}$ and
where the approximation in Eq.~(\ref{Lmix-grad}) holds up to $O({\bf Z}^{-3})$.
In lowest order in $1/Z_i$, we find, 
with $\varphi_{12}=(\bphi_{01}\cdot \bar{\bf m}_1)(\bphi_{01}\cdot \bar{\bf m}_2)$,
\begin{equation}
  \frac{1}{T_1^{\rm (mix)}} = -\!\left(\!\frac{\Phi_0}{2\pi}\!\right)^2 
\!\!\!{\rm Im}\frac{8\varphi_{12}\omega_{01}^2 L_{12} }{Z_1(\omega_{01}) Z_2(\omega_{01})}  
\coth\left(\!\frac{\beta\omega_{01}}{2}\!\right)\!\!.
\end{equation}

If $R_i\equiv Z_i(\omega_{01})$ are real (pure resistances) then, as predicted above,
the imaginary part of the second-order term
in Eq.~(\ref{Lmix-grad}) vanishes, and we resort to third order,
\begin{equation}
  {\rm Im}{\bf L}_{\rm mix}^{-1} = \frac{\omega^3 L_{12}}{R_1 R_2}\left(\begin{array}{c c}
                               \frac{L_{12}}{R_1} & \frac{L_{11}}{R_1}+\frac{L_{22}}{R_2}\\
                               \frac{L_{11}}{R_1}+\frac{L_{22}}{R_2} & \frac{L_{12}}{R_2}
                           \end{array}\right),
\end{equation}
neglecting terms in $O(R_j^{-4})$.
If $L_{12}\ll L_{jj}$, we obtain
${\rm Im}{\bf L}_{\rm mix}^{-1} \approx 
(\omega^3 L_{12}/R_1 R_2) (L_{11}/R_1+L_{22}/R_2)\sigma_x$,
and
\begin{equation}
  \frac{1}{T_1^{\rm (mix)}} = \left(\frac{\Phi_0}{2\pi}\right)^2\frac{8\omega_{01}^3 L_{12}}{R_1 R_2} \left(\frac{L_{11}}{R_1}\!+\!\frac{L_{22}}{R_2}\right)\varphi_{12} \coth\left(\frac{\beta\omega_{01}}{2}\right).
\end{equation}

For the gradiometer qubit (Fig.~\ref{circuit}), we find
$L_{12}\approx M_{12}M_{13}M_{34}/L_1 L_3$, $L_{11}\approx L_2$, $L_{22}\approx L_4$,
where $L_k$ denotes the self-inductance of branch $X_k$ ($X$=$L$ or $K$) and $M_{kl}$
is the mutual inductance between branches $X_k$ and $X_l$,
and where we assume $M_{ij} \ll L_k$.
The ratio between the mixing the single-impedance contribution scales as
\begin{equation}
  \frac{1/T_1^{{({\rm mix})}}}{1/T_1^{(j)}} 
       \approx \frac{\omega_{01}^2 L_{12} L}{R^2},
\end{equation}
where we have assumed $R_1\approx R_2 \equiv R$, $L_{11}\approx L_{22} \equiv L$,
and $\bphi_{01}\cdot \bar{\bf m}_1 \approx \bphi_{01}\cdot \bar{\bf m}_2$.

We have calculated $T_1$ at temperature $T \ll \hbar\omega_{01}/k_B$ for the circuit Fig.~\ref{circuit},
for a critical current $I_c=0.3\,\mu {\rm A}$ for all junctions, and for the
inductances $L_1=30\,{\rm pH}$, $L_3=680\,{\rm pH}$, $L_2=L_4=12\,{\rm nH}$,
$M_{12}\simeq \sqrt{L_1 L_2}$, $M_{34}\simeq \sqrt{L_3 L_4}$ (strong inductive coupling),
$M_{35}=6\,{\rm pH}$, with $\omega_{01}=2\pi\cdot 30\,{\rm GHz}$, 
and with the impedances $Z_1=R$, $Z_2=R + i R_{\rm im}$, where $R$ and $R_{\rm im}=\pm 10\,{\rm k}\Omega$
are real ($R_{\rm im}>0$ corresponds to an inductive, $R_{\rm im}<0$ to a capacitive character of $Z_i$).  
In Fig.~\ref{fig2}, we plot $T_1$ with and without mixing for a fixed value of $M_{13}=0.5\,{\rm pH}$
and a range of $R={\rm Re}Z_i$.
In the inset of Fig.~\ref{fig2}, we plot $T_1$ (with mixing) and 
$((T_1^{(1)})^{-1}+(T_1^{(2)})^{-1})^{-1}$ (without mixing) for $R=75\,\Omega$ for a range
of mutual inductances $M_{13}$;  for this plot, we numerically computed the double minima of the 
potential $U$ and $\bphi_{01}$ for each value of $M_{13}$.
The plots (Fig.~\ref{fig2}) clearly show that summing the decoherence rates without taking into
account mixing term can both underestimate or overestimate the relaxation rate $1/T_1$, leading 
to either an over- or underestimate of the relaxation and decoherence times $T_1$ and $T_2$.

\textit{Higher-order terms in the Born series.}
Two series expansions have been made in our analysis, (i) the Born approximation to lowest order
in the parameter $\alpha_B \approx \mu R_Q/Z_i(\omega_{01}) \approx 1/\omega_{01} T_1$, 
where $\mu$ is a dimensionless ratio of inductances \cite{BKD} and $R_Q=h/e^2$ is the quantum of resistance,
and (ii) the expansion Eq.~(\ref{series1}) in the parameter $\alpha_L \approx \omega_{01} L/Z_i$, 
where $L$ is the inductance of the circuit, where we included higher orders.
The question arises whether the terms in the next order in $\alpha_B$ in the Born approximation could be of
comparable magnitude to those taken into account in $1/T_1^{({\rm mix})}$.
In our example, we could neglect such terms, because $\alpha_B / \alpha_L \approx 0.001/0.1 = 0.01 \ll 1$, 
but in cases where $\alpha_B \gtrsim \alpha_L$, higher orders
in the Born approximation may have to be taken into account.

\textit{Acknowledgments.}
We thank David DiVincenzo and Roger Koch for useful discussions.
FB would like to acknowledge the hospitality of the
Quantum Condensed Matter Theory group at Boston University.
FB is supported by Funda\c{c}\~{a}o da Amparo \`{a} Pesquisa do
Estado de S\~{a}o Paulo (FAPESP).

\end{document}